\documentclass[12pt]{article}

\font\bfs=cmb10 at 7pt

\font\cal=cmsy10 at 9pt
\font\cals=cmsy10 at 8pt

\font\rms=cmr10 at 8pt

\font\sf=cmss10 at 10pt 

\def\0#1{\mbox{\rm#1}}
\def\1#1{\mbox{\bb#1}}
\def\2#1{\mbox{\bf#1}}
\def\3#1{{\cal #1}}
\def\4#1{\mbox{\cals#1}}
\def\5#1{\mbox{\sf#1}}
\def\6#1{\mbox{\rms #1}}
\def\7#1{\mbox{\bfs #1}}
\def\8#1{{\tilde #1}}
\def\9#1{{\breve #1}}

\def\BEq{\begin{equation}}
\def\EEq{\end{equation}}
\def\BEqA{\begin{eqnarray}}
\def\EEqA{\end{eqnarray}}
\def\BEn{\begin{enumerate}}
\def\EEn{\end{enumerate}}

\def\tav{\hbox{
\kern-1pt\rule[0pt]{1.5pt}{.8pt}{\kern-3.3pt}
\rule[0pt]{.4pt}{5pt}{\kern-4.4pt}
\rule[5pt]{4.5pt}{.8pt}{\kern-3.45pt}
\rule[0pt]{.4pt}{5.3pt}{\kern-1pt}
}}

\def\a{\alpha}

\def\d{\delta}

\def\e{\epsilon}

\def\m{\mu}

\def\n{\nu}

\def\t{\tau}

\def\from{\kern-2pt\leftarrow\kern-2pt}

\def\II{|\kern-1pt |}

\def\tav{\hbox{
\kern-1.0pt
\rule[0pt]{1.3pt}{.8pt}{\kern-3.6pt}
\rule[0pt]{.4pt}{6pt}{\kern-3.0pt}
\rule[4.5pt]{3.0pt}{.8pt}{\kern-3.3pt}
\rule[0pt]{.4pt}{5pt}{\kern-1pt}
}}

\def\from{\kern-2pt\leftarrow\kern-2pt}

\def\lmult{{\lfloor\kern-5pt\lfloor}}
\def\rmult{{\rfloor\kern-5pt\rfloor}}

\def\II{|\kern-1pt |}

\usepackage{makeidx,graphics,latexsym}

\begin{document}

\title{{\bf Tunneling out of metastable vacuum in a system consisting of two capacitively coupled phase qubits}}

\author{
Andrei
Galiautdinov
\\
\normalsize{\it Brenau University, Gainesville, Georgia
30501, U.S.A.}
}


\maketitle

\abstract{Using a powerful combination of Coleman's instanton technique and the method of Banks and Bender, the exponential factor for the zero temperature rate of tunneling out of metastable vacuum in a system of two identical capacitively coupled phase qubits is calculated in closed form to second order in asymmetry parameter for a special case of intermediate coupling $C=C_J/2$.}

\tableofcontents



\section{Introduction}

Tony Leggett's beautiful insights \cite{LEGGETT80, CALDEIRALEGGETT83} into the nature of macroscopic quantum phenomena have long been an inspiration to many students of quantum mechanics. One particularly interesting system whose macroscopic variables exhibit quantum mechanical behavior predicted by Leggett is the current biased Josephson junction (CBJJ). Since the mid 80's a considerable amount of experimental and theoretical work has been done to understand its properties \cite{CLARKEETAL88}. It was eventually realized that due to its characteristic energy spectrum a CBJJ may play an important role in the field of quantum computation \cite{MAKHLIN01}. The lowest quasistationary energy levels of the junction, whose spacings are relatively easy to manipulate, may be viewed as the levels of a qubit, and if several such qubits are joined together a quantum computer could be built and operated.

The physics of an isolated junction is by now well understood. However, when two or more such junctions are coupled together the calculations of their properties become much more involved. This is because we have to deal with higher dimensional problems. The matter is further complicated by the presence of additional velocity dependent terms in the corresponding Lagrangians. 

One important property that needs to be addressed is the rate of tunneling out of metastable vacuum in a system of several junctions. Such tunneling process (appearance of a voltage drop across the junctions) is often used to perform a readout on qubits, and therefore a good theoretical understanding of it is essential to the general understanding of CBJJ-based quantum computers. 

In this paper we propose a method for calculating the {\it zero temperature} tunneling {\it exponential factor}. It combines Coleman's instanton technique \cite{COLEMAN-I} and the approach of Banks and Bender \cite{BANKSBENDER73} based on Jacobi's variational procedure.

With respect to the proposed method the following must be kept in mind:

\begin{enumerate}
\item The tunneling rate is calculated using Coleman's formula (\ref{eq:COLEMANSFORMULA}). The action is for the path of {\it zero energy} which starts at the top of the {\it inverted} potential $U_E$ (corresponding to the metastable minimum of $U$), approaches the turning point of $U_E$ (the exit point of $U$), then goes back to the top. Such path is called a bounce.

\item The action of the system is stationary under variations of the bounce. Being a path of zero energy, the bounce is therefore the most probable escape path for tunneling out of metastable vacuum. This indicates that the results obtained here should be in agreement with those derived using the WKB approximation.

\item In the presence of velocity dependent couplings, even for highly symmetrical potentials (double wells, etc.) the bounce may not be a straight line. In general its shape depends on the interplay between the various kinetic energy terms.

\item Because the action appears in the exponent, the calculated tunneling rate strongly depends on how much the bounce actually curves.

\item In our problem, which deals with identical qubits, only one metastable well is considered. The relevant bounce at zero bias difference is indeed a straight segment despite the presence of a velocity dependent term in the Lagrangian.

\end{enumerate}

\section{Lagrangian for capacitively coupled qubits}

We are interested in a system of two phase qubits coupled by a capacitor \cite{JOHNSONETAL2003} as depicted on Figure 1.
The bias currents are denoted by $I_1$ and $I_2$, $C_{1,2}$ are junctions' effective capacitances, $C$ is the coupling.
Considering the balance of currents in the circuit
\BEq
 I_1=I_{J_1}+I_{C_1}+I_{C}, \quad
 I_2=I_{J_2}+I_{C_2}-I_C,
\EEq
and using Josephson's equations for the voltage and phase difference across a junction whose critical current is $I_0$,
\BEq
I_J=I_0 \sin \phi, \quad \dot \phi = \a V,
\EEq
we are led to the equations of motion for the two superconducting phases
\BEqA
         \a^2[(C_1+C)\ddot{\phi_1}-C \ddot{\phi_2}]+\a(I_{01}\sin \phi_1 -I_1)&=&0\;,\\
         \a^2[-C \ddot{\phi_1}+(C_2+C)\ddot{\phi_2}]+\a(I_{02}\sin \phi_2 -I_2)&=&0\;,
\EEqA
where $\a \equiv \hbar /2e $. The corresponding Lagrangian is given by
\BEq
L=\a^2\left[ \frac{(C_1+C)\dot{\phi_1}^2}{2} 
+ \frac{(C_2+C)\dot{\phi_2}^2}{2} - C\dot\phi_1 \dot\phi_2
\right]
-U(\phi_1, \phi_2; I_1, I_2),
\EEq
where
\BEq
U(\phi_1, \phi_2; I_1, I_2)=
- \a\left(I_{01}\cos \phi_1+I_1\phi_1+I_{02}\cos \phi_2+I_2\phi_2\right),
\EEq
and the dot $ \dot{} $ represents differentiation w.r.t. the time $t$.

Denote $j_i\equiv I_i/I_{0i}$. As $j_{1,2}\rightarrow 1^-$, $\phi_{1,2}\rightarrow \pi/2$. In this limit,
\BEqA
U &\simeq&  -\left( \epsilon_1 j_1+\e_2 j_2\right)\pi/2 
+ \;  \e_1(1-j_1)\left( \phi_1 -\pi/2\right)
+\e_2(1-j_2)\left(\phi_2-\pi/2\right)
\nonumber \\
&& \quad \quad \quad
- \; \e_1/6 \left(\phi_1-\pi/2\right)^3
- \e_2/6\left(\phi_2-\pi/2\right)^3 \,,
\EEqA
where $\e_i\equiv \a I_{0i}$ are junctions' Josephson energies. Shifting the local energy minimum to the origin gives
\BEqA
U &=& \frac{1}{6}
\left[
\e_1 \phi_1^2 \left( a-\phi_1\right) 
+
\e_2 \phi_2^2 \left( b-\phi_2\right) 
\right]
\nonumber \\
& \equiv &  \frac{27}{4}
\left[
\Delta U_1 \left(\frac{\phi_1}{a}\right)^2 
\left( 1-\frac{\phi_1}{a}\right)
+
\Delta U_2 \left(\frac{\phi_2}{b}\right)^2 
\left( 1-\frac{\phi_2}{b}\right)
\right]\,,
\EEqA
where
\BEq
a=3\sqrt{2(1-j_1)}, \quad b=3\sqrt{2(1-j_2)}
\EEq
are the exit points of the potential in the directions of the coordinate axes and
\BEq
\Delta  U_i = \frac{4\sqrt{2}}{3}\e_i(1-j_i)^{3/2}
\EEq
are the corresponding potential heights. 

The use of instanton technique \cite{COLEMAN-I} for tunneling rate calculations requires the introduction of the Euclidean Lagrangian and equations of motion (in non-physical time $\t$), which in this paper will be used interchangeably with the usual ones.

Introducing a small asymmetry parameter $\eta $ such that $b\equiv (1+\eta)a$, the full Euclidean Lagrangian becomes
\BEq
\label{eq:LAGRANGIANVELOCITYDEPENDENT}
L_E=\frac{m_1 \dot{\phi_1}^2}{2} 
+ \frac{m_2 \dot{\phi_2}^2}{2} - m_3 \dot \phi_1 \dot \phi_2
- U_E(\phi_1,\phi_2; \eta)\;,
\EEq
where 
\BEq
 U_E(\phi_1,\phi_2; \eta)\equiv -U(\phi_1,\phi_2; \eta)=\frac{1}{6}
\left[\e_1 \phi_1^3+\e_2\phi_2^3-a(\e_1\phi_1^2+\e_2\phi_2^2)-\eta a \e_2\phi_2^2
\right],
\EEq
$m_i \equiv \a^2(C_i+C)$, and $m_3 \equiv \a^2 C$.

For identical qubits, $C_1=C_2\equiv C_J$, $I_{01}=I_{02}\equiv I_0$, $m_1=m_2\equiv m+m_3 = \a^2(C_J+C)$, $m=\a^2C_J$, $\e_1=\e_2\equiv \e$, so that
\BEq
\label{eq:UEXY}
 L_E(\phi_1, \phi_2; \eta)=\frac{m+m_3}{2} \left(\dot{\phi_1}^2+\dot{\phi_2}^2\right) - m_3 \dot \phi_1 \dot \phi_2-\frac{\e}{6}
\left[\phi_1^3+\phi_2^3-a(\phi_1^2+\phi_2^2)-\eta a \phi_2^2 \right].
\EEq
We will also need an expression for $L_E$ in the ``diagonal'' coordinates
\BEq
x_1= \frac{1}{\sqrt{2}}\left(\phi_1+\phi_2\right), \quad
x_2= \frac{1}{\sqrt{2}}\left(\phi_1-\phi_2\right),
\EEq
which is
\BEq
\label{eq:UEX1X2}
 L_E(x_1,x_2;\eta)=\frac{m\dot{x_1}^2}{2} 
+ \frac{M \dot{x_2}^2}{2}-\frac{\e}{6}
\left[\frac{1}{\sqrt{2}}\left(x_1^3+3x_1 x_2^2\right)-a\left(x_1^2+x_2^2\right)-\frac{1}{2}\eta a (x_1-x_2)^2\right],
\EEq
where $M\equiv m+2m_3=\a^2(C_J+2C)$. The three-dimensional graphs of the potential in symmetric ($\eta = 0$) and asymmetric ($\eta = 0.30$) cases are given on Figures 2 and 3. Here, $m=1$ and $\e =1$. In the $x_{1,2}$-coordinates
the critical points of $U_E$ are $(0, \,0)$ -- local maximum, $ \left(\sqrt{2}a/3, \, \sqrt{2}a/3\right)$ and $\left((1+\eta)\sqrt{2}a/3, \, -(1+\eta )\sqrt{2}a/3\right)$ -- saddle points, and $\left((2+\eta)\sqrt{2}a/3, \, -\eta \sqrt{2}a/3 \right)$ -- local minimum. Figures 4 and 5 demonstrate the locations of these points on the corresponding contour diagrams in the symmetric and asymmetric cases.

It is important to clearly understand the meaning of the asymmetry parameter $\eta$. It is due entirely to the difference in the bias currents $j_i$. The reference current is arbitrarily chosen to be $j_1$. Experiment begins with both bias currents pre-set at some $j_{\rm initial}=j_1$. This corresponds to a given value of the exit point $a$ in the $\phi_1$ direction. Any change in the tunneling rate is always measured in relation to the $a$ so chosen. Varying $j_2$ changes the effective width and the height of the potential barrier in the $\phi_2$ direction. An {\it increase} $\Delta j$ in $j_2$ effectively {\it reduces} the size of the barrier, resulting in a {\it negative} value of $\eta$, which is explicitly given by
\BEq
\label{eq:ETA}
\eta = \sqrt{1-\frac{\Delta j}{1-j_{\rm initial}}}-1.
\EEq
That, in turn, leads to {\it higher} tunneling rate as will be rigorously demonstrated below.
 
\section{A note on instantons}

In the {\it symmetric case} ($\eta =0$), the equations of motion derived from (\ref{eq:UEXY}) and (\ref{eq:UEX1X2}) using the usual variational procedure are
\BEqA
\label{eq:EQXY}
(m+m_3)\ddot \phi_1-m_3 \ddot \phi_2 &=&-\frac{\e}{6}\left(3\phi_1^2-2a\phi_1\right)\,,
\nonumber \\
(m+m_3)\ddot \phi_2-m_3 \ddot \phi_1 &=&-\frac{\e}{6}\left(3\phi_2^2-2a\phi_2\right)\,,
\EEqA
and
\BEqA
\label{eq:EQX1X2}
\ddot x_1&=&-\frac{\e}{6m}\left[\frac{3}{\sqrt{2}} (x_1^2+x_2^2)-2ax_1\right]\,,
\nonumber \\
\ddot x_2&=&-\frac{\e}{6M}\left[\frac{6}{\sqrt{2}} x_1x_2-2ax_2\right]\,.
\EEqA
We now ask: What is the effect of coupling $C\sim m_3$ on the {\it instanton} solutions of (\ref{eq:EQXY}) satisfying the boundary and energy $(E=0)$ conditions
\BEq
\label{eq:BOUNDARYCONDS}
\vec{\phi}(\pm \infty)=0, \;\dot{\vec{\phi}}(0)=0?
\EEq

When $C=0$ (decoupled qubits), there are {\it three} such solutions:
\begin{enumerate}
\item one passing through the saddle point of $U_E=-U$ in the $\phi_1$ direction,
\BEq
\label{eq:INSTANTONX}
\phi_1^{(0)}(\t)=\frac{a}{\cosh^2(\omega \t/2)}, \quad \phi_2^{(0)}(\t)=0,
\EEq
\item another passing through the saddle point in the $\phi_2$ direction,
\BEq
\label{eq:INSTANTONY}
\phi_1^{(0)}(\t)=0, \quad \phi_2^{(0)}(\t)=\frac{a}{\cosh^2(\omega \t/2)},
\EEq
and
\item the third one in the $x_1 \sim \phi_1+\phi_2$ direction of $U_E$'s local minimum ($U$'s local maximum),
\BEq
\label{eq:INSTANTONX1}
x^{(0)}_1(\t)=\frac{a\, \sqrt{2}}{\cosh^2(\omega \t/2)}, \quad x^{(0)}_2(\t)=0.
\EEq
\end{enumerate}
Here, the superscript ${}^{(0)}$ reminds us that $\eta = 0$. Also,
\BEq
\label{eq:ATTEMPTFREQSYM}
\omega = \sqrt{\frac{1}{m}\frac{\partial^2 U}{\partial x_1^2}}=\sqrt{\frac{\e a}{3m}}
\EEq
is the frequency of small normal oscillations in the $x_1$ direction. 
This frequency is also called the attempt frequency because it appears in the tunneling rate prefactor. It can be shown by direct calculation that for a slightly asymmetric $U$ the attempt frequency acquires a linear in $\eta$ contribution, $\omega \rightarrow \omega (1+\eta/4)$.

Notice that {\it in configuration space} all three instantons are straight segments. Also notice that (\ref{eq:INSTANTONX1}) should be regarded as non-physical since the two qubits are completely decoupled in this case \cite{GELLERcomment}.

When $C > 0$, there are no longer instanton solutions in the form of straight segments passing through the saddle points. Indeed, if, for example, we choose to look for a solution in the form
\BEq
\phi_1=0, \quad \phi_2=\phi_2(\t),
\EEq
we would get the system
\BEqA
-m_3 \ddot \phi_2 &=&0, \nonumber \\
(m+m_3)\ddot \phi_2 &=&-\frac{\e}{6}\left(3\phi_2^2-2a\phi_2\right),
\EEqA
with two constant solutions $\phi_2=0$ and $\phi_2=2a/3$, which are {\it not} instantons. So (\ref{eq:INSTANTONX}) and (\ref{eq:INSTANTONY}) do change when $C>0$.

In this regard we notice that, according to (\ref{eq:EQXY}) and (\ref{eq:BOUNDARYCONDS}), in the extreme case of very strong coupling ($C\gg C_J$), $\ddot \phi_1(\t) -\ddot \phi_2(\t) \approx 0 \Rightarrow \phi_1(\t)-\phi_2(\t)\approx At+B \Rightarrow \phi_1(\t)\approx \phi_2(\t)$, so both instantons disappear.
Apparently the velocity dependent coupling becomes so strong that it prevents the formation of the corresponding bounces.

However the $x_1$-instanton described by (\ref{eq:INSTANTONX1}) becomes physical, surviving any coupling. Any change in it is due to asymmetry in the potential.

We now come to a subtle point: {\it Do the $\phi_1$- and $\phi_2$-bounces  actually exist at finite coupling $C >0$}?

Here is an argument I would like to put forward regarding this matter: Assume all three exist. Then the tunneling rate is roughly the sum of three contributions,
$e^{-S_{\phi_1}/\hbar}+e^{-S_{\phi_2}/\hbar}+e^{-S_{x_1}\hbar}$,
where various $S$'s are the actions for the corresponding instantons.
Now, as $C \rightarrow \infty$, $P\rightarrow e^{-S_{x_1}/\hbar}$, which is equivalent to 
\BEq
\label{eq:SXYLIMIT}
S_{\phi_1}, \, S_{\phi_2} \sim \int \sqrt{2U} \rightarrow \infty.
\EEq
 However, as we have just shown, in this limit the bounces must satisfy the condition $\phi_1(\t)\approx \phi_2(\t)$. This contradicts (\ref{eq:SXYLIMIT}) because in the $\phi_1\approx \phi_2>0$ region $U$ is continuous and bounded. We therefore conclude that {\it at any finite coupling there is one and only one instanton} given by (\ref{eq:INSTANTONX1}).
 
 The physical meaning of this conclusion is the following: When two qubits are capacitively coupled there can be no ``tunneling'' voltage drop on one of them without having a comparable drop on the other. Apparently the qubits always ``tunnel'' together. This is especially obvious in the limit $C\rightarrow \infty$ when the qubits effectively become one.

\section{Jacobi's variational principle}

We now turn to the effect of asymmetry ($\eta \neq 0$) on the instanton solution (\ref{eq:INSTANTONX1}). It is convenient to work with the
equations of motion (with energy condition $E=0$) derived via Jacobi's variational principle
\BEq
\d \int_1^2 ds \sqrt{2U(s)}=0
\EEq
with the constraint
\BEq
\label{eq:CONSTRAINT}
ds^2=m (dx_1)^2+M (dx_2)^2.
\EEq
Performing the variation we get ({\it cf.} \cite{BANKSBENDER73})
\BEqA
\label{eq:EQMOTIONs}
2Um \frac{d^2x_1}{ds^2}+m\frac{dx_1}{ds}
\left( \frac{dx_1}{ds}\frac{\partial U}{\partial x_1}
+\frac{dx_2}{ds}\frac{\partial U}{\partial x_2}\right)
&=& \frac{\partial U}{\partial x_1}\,,
\nonumber \\
2UM\frac{d^2x_2}{ds^2}+M\frac{dx_2}{ds}
\left( \frac{dx_1}{ds}\frac{\partial U}{\partial x_1}
+\frac{dx_2}{ds}\frac{\partial U}{\partial x_2}\right)
&=& \frac{\partial U}{\partial x_2}\,.
\EEqA
Let us assume that these equations have been solved in the symmetric case $(\eta=0)$: ${\bf x }^{(0)}(s)=(x_1^{(0)}(s), x_2^{(0)}(s))$. The solution in the asymmetric case $(\eta \neq 0)$ will be written in the form of a power series in $\eta$:
\BEq
{\bf x }(s)={\bf x }^{(0)}(s)+\eta {\bf x }^{(1)}(s)+\eta^2 {\bf x }^{(2)}(s)+\dots  .
\EEq

From (\ref{eq:CONSTRAINT}) several important relations among various ${\bf x }^{(k)}(s)$ can be derived. [From here on, the dot $\dot{}$ stands for the derivative with respect to $s$ and the dot product is evaluated with the help of the metric ${\rm diag}(m, M)$.] To second order in $\eta$ we have:
\BEqA
1=\dot{\bf x }\cdot \dot{\bf x }&=& 
\left( \dot{\bf x }^{(0)} +\eta \dot{\bf x }^{(1)} +\eta^2 \dot{\bf x }^{(2)} \right)\cdot \left( \dot{\bf x }^{(0)}+\eta \dot{\bf x }^{(1)}+\eta^2 \dot{\bf x }^{(2)}\right) \nonumber \\
&=& 1 + 2 \eta \, \dot{\bf x }^{(0)}\cdot \dot{\bf x }^{(1)}+
\eta^2 \left(2 \dot{\bf x }^{(0)}\cdot \dot{\bf x }^{(2)}
+\dot{\bf x }^{(1)} \cdot \dot{\bf x }^{(1)} \right).
\EEqA
 This leads to
\BEq
\label{eq:CONSTR1}
\dot{\bf x }^{(0)}\cdot \dot{\bf x }^{(1)}=0
\EEq
and
\BEq
\label{eq:CONSTR2}
\dot{\bf x }^{(0)}\cdot \dot{\bf x }^{(2)}
=-\frac{1}{2}\dot{\bf x }^{(1)} \cdot \dot{\bf x }^{(1)} .
\EEq

We will take ${\bf x }^{(0)}(s)$ to be the $s$-parametrized form of the instanton solution ${\bf x }^{(0)}(\t)$ given by (\ref{eq:INSTANTONX1}). Thus in configuration space ${\bf x }^{(0)}(s)$ is a straight segment
\BEq
\label{eq:PARAMETRIZEDSOLITON1stORDER}
{\bf x}^{(0)}(s)
 = s \, \left(1/\sqrt{m},\, 0\right), \quad 0\leq s \leq s_1, \quad s_1=a\, \sqrt{2m}.
\EEq

Now, from (\ref{eq:CONSTR1}) and (\ref{eq:PARAMETRIZEDSOLITON1stORDER})
it follows that the first component of $\dot{\bf x}^{(1)}(s)$ must be zero which allows us to choose ${\bf x}^{(1)}(s)$ in the form
\BEq
\label{eq:CHI}
{\bf x}^{(1)}(s)=(0, \chi (s)).
\EEq
Thus to {\it second order in $\eta$},
\BEq
{\bf x}(s)\equiv \left( x_1(s), \,  x_2(s)\right) 
=\left( x_1^{(0)}(s)+\eta^2 x_1^{(2)}(s) , \;  \eta \chi(s) + \eta^2 x_2^{(2)}(s)\right),
\EEq
and the potential is given by
\BEqA
\label{eq:UX1X2SECONDORDER}
 -6U(\eta)/\e&=&
\left(x_1^{(0)}\right)^2
\left( \frac{1}{\sqrt{2}}\, x_1^{(0)}-a \right)
\nonumber \\
&& \; -\; \eta \, \frac{a}{2}\left(x_1^{(0)}\right)^2
 + \eta^2 
\left[
\frac{3}{\sqrt{2}}\, x_1^{(0)}\left(x_1^{(0)}x_1^{(2)}+\chi^2\right)
+a\left(x_1^{(0)}\chi -2x_1^{(0)}x_1^{(2)}-\chi^2\right)
 \right]
\nonumber \\
&=&
\frac{s^2}{m\sqrt{2m}}
\left( s-s_1 \right)
\nonumber \\
&& -\;\frac{\eta s^2}{m\sqrt{2m}}\frac{s_1}{2}
+\frac{\eta^2 }{\sqrt{2m}}
\left[
3s\left(\frac{s \, x_1^{(2)}}{\sqrt{m}}+\chi^2\right)
+s_1
\left(\frac{s\left( \chi -2x_1^{(2)}\right)}{\sqrt{m}}-\chi^2\right)
 \right].
\EEqA
From (\ref{eq:CONSTR2}), (\ref{eq:PARAMETRIZEDSOLITON1stORDER}), and (\ref{eq:CHI}) it follows that 
\BEq
\dot x_1^{(2)}
=-\frac{1}{2}\frac{M}{\sqrt{m}} \, \dot{\chi}^2 \,,
\EEq
or
\BEq
x_1^{(2)}(s)
=-\frac{1}{2}\frac{M}{\sqrt{m}} \int_0^s d\tilde{s} \; \dot{\chi}^2(\tilde{s}) \,.
\EEq
We see that, as pointed out in \cite{BANKSBENDER73}, the calculation of {\it tunneling effects to second order in $\eta$} requires the knowledge of the {\it tunneling path to only first order in $\eta$}.

Thus, to first order,
\BEqA
\frac{\partial U(s)}{\partial x_1}&=&-\frac{\e s}{6\sqrt{2}m}\left(3s-2s_1-\eta s_1\right),
\\
\frac{\partial U(s)}{\partial x_2}&=&-\frac{\e \eta}{6\sqrt{2m}} \left(2(3s-s_1)\chi+\frac{ss_1}{\sqrt{m}}\right),
\EEqA
and from (\ref{eq:EQMOTIONs}) we get the following equation for $f(s) \equiv \sqrt{m}\chi(s)$:
\BEq
\label{eq:EQfs}
2s^2(1-s/s_1)\ddot f(s) + s(2-3s/s_1)\dot f(s) - 2\kappa (1-3s/s_1)f(s) 
= -\kappa s,
\EEq
where
\BEq
 \kappa \equiv m/M = C_J/(C_J+2C)
\EEq
characterizes the strength of the capacitive coupling.

We may further perform the change of variables
\BEq
f(s)\rightarrow g(z)=f(z)/s_1, \quad s \rightarrow z=\left(1-s/s_1\right)^{1/2},
\EEq
with $z$ ranging from 1 to 0, transforming (\ref{eq:EQfs}) into
\BEq
\label{eq:LEGENDRE}
\left(1-z^2\right)\frac{d^2g(z)}{dz^2}-2z\frac{dg(z)}{dz}+\kappa \left[12-\frac{4}{1-z^2}\right]g(z)=-2\kappa .
\EEq
This we recognize as an associated inhomogeneous Legendre equation whose homogeneous solutions are associated Legendre functions $P_{\n}^{\m}$ and $Q_{\n}^{\m}$ of degree $\n$ (where $\n (\n +1)=12\kappa$) and order $\m=2\sqrt{\kappa}$\,. 

\section{The case of intermediate coupling $C=C_J/2$}

Up to this point the derivation was valid for any coupling. However equation (\ref{eq:LEGENDRE}) is difficult to work with in general. To simplify the presentation we consider a special case $\kappa = 1/2$ which corresponds to {\it intermediate coupling} $C=C_J/2$,
\BEq
\label{eq:LEGENDRESPECIALCASE1/2}
\left(1-z^2\right)\frac{d^2g(z)}{dz^2}-2z\frac{dg(z)}{dz}+ \left[6-\frac{2}{1-z^2}\right]g(z)=-1 .
\EEq
The solution to (\ref{eq:LEGENDRESPECIALCASE1/2}) satisfying the all-important initial condition $g(1)=0$ is 
\BEq
g(z)=(z^2-1)/2
\EEq
which gives
\BEq
\chi(s)=-\frac{s}{2\sqrt{m}}, \quad \chi(0)=0.
\EEq
This leads to a very simple expression for the potential $U$ in (\ref{eq:UX1X2SECONDORDER}),
\BEq
\label{eq:UX1X2SECONDORDERINTERMEDIATECASE}
U(\eta)=
\frac{\e s^2}{6m\sqrt{2m}}
\left( s_{\rm exit}-s \right),
\EEq
where
\BEq
\label{eq:EXITPOINT}
s_{\rm exit}=s_1 \left(  1+\frac{1}{2}\eta+\frac{1}{4}\eta^2\right)
\EEq
is its exit point.

\section{Results of numerical simulations}

A Runge-Kutta simulation of the $x_1$-instanton trajectory in configuration space is presented in the following table (where, again, $\eta = 0.30$, $m=1$, and $\e =1$):
\begin{equation}
 \begin{array}{|c|c|c|c|c|}\hline
 s & x^{\rm sim.}_1 & x^{\rm pred.}_1 & x^{\rm sim.}_2 & x^{\rm pred.}_2 \\ \hline
0.000 & 0.000& 0.000& -0.000& -0.000 \\ \hline
0.500 & 0.491 & 0.489 & -0.066 & -0.075 \\ \hline
 1.000 & 0.983 & 0.978 & -0.130 & -0.150 \\ \hline
 1.500 & 1.475 &1.466 & -0.193 &  -0.225 \\ \hline
 1.654 & 1.626 &1.617 & -0.213 &  -0.248 \\ \hline
 \end{array}
\end{equation}
The predicted values of $x^{\rm pred.}_1$ and $x^{\rm pred.}_2$ are given to second and first order in $\eta$, respectively, since that is the accuracy to which they are needed for the second order tunneling rate calculations.

The corresponding exit point has also been found numerically. It is $s^{\rm sim.}_{\rm exit}=1.654$, which is in good agreement with the value $s^{\rm pred.}_{\rm exit}=1.658$ predicted using (\ref{eq:EXITPOINT}) .

\section{The exponential factor}

The tunneling exponential factor $P$ can now be calculated \cite{COLEMAN-I} via
\BEq
\label{eq:COLEMANSFORMULA}
P=\exp{\left(-S_0/\hbar \right)},
\EEq
where
\BEq
S_0=2\int_0^{s_{\rm exit}} ds \, \sqrt{2U(s)}
\EEq
is the action for the tunneling path ${\bf x}(s)$. The factor of 2 is due to the bounce.

Using (\ref{eq:UX1X2SECONDORDERINTERMEDIATECASE}) we find:
\BEqA
S_0&=&2\; \sqrt{ \frac{\epsilon}{ 3m\sqrt{2m} }  }\,\int_0^{s_{\rm exit}}ds \; s \, \sqrt{s_{\rm exit}-s}
\nonumber \\
&=& 2\; \sqrt{ \frac{\epsilon}{ 3m\sqrt{2m} }  }\,
\frac{4}{15} \, (s_{\rm exit})^{5/2}
\nonumber \\
&\equiv & S_0^{(0)}\left[1+\frac{5}{4}\, \eta+\frac{35}{32}\, \eta^2\right], \\ \nonumber \\
S_0^{(0)}&= &\frac{16}{15}\, m \omega a^2 =
\frac{96}{5}\,
\sqrt{
\left(\frac{\hbar}{2e}\right)^3C_JI_0
\sqrt{2\left(1-j_{\rm initial}\right)^5}
},
\EEqA
where the terms to second order have been retained and $\eta$ is given by (\ref{eq:ETA}).

Notice that the zeroth order term (corresponding to $\kappa =0$, $\eta =0$; strongly coupled qubits, zero bias difference) coincides with the well known result of Caldeira and Leggett \cite{CALDEIRALEGGETT83} (for instanton of amplitude $a\sqrt{2}$). Also, negative $\eta$ (higher bias) increases the tunneling rate, as expected.

\section{Conclusions and the outlook}

In this paper a combination of instanton technique and Jacobi's variational procedure has been used for calculation of the rate of tunneling out of metastable vacuum in a system consisting of two capacitively coupled phase qubits. In the interesting case of intermediate coupling the tunneling rate was explicitly calculated in closed form to second order in asymmetry parameter $\eta$.

As a by-product, a detailed analysis of instanton trajectories in the presence of velocity dependent coupling has been performed. It was found that a curved version of only one such trajectory, (\ref{eq:INSTANTONX1}), gives the major contribution to the tunneling rate in a capacitively coupled system. Physically this means that simultaneous tunneling of both qubits is overwhelmingly more probable than any other mode of tunneling.

Tunneling out of higher energy levels and tunneling out of various superpositions of the ground and excited states is an active area of modern experimental research. While the methods of this paper can be easily applied to tunneling out of metastable vacuum, it is important to know whether more complicated tunneling processes can be analysed using the same approach.

The author believes that generalization to higher energy states is straightforward. Again, the most probable escape path would be given by a curved version of the bounce along the $x_1$ axis, similar to (\ref{eq:INSTANTONX1}). This time, however, the bounce would pass through the exit points of $U_E$ corresponding to non-zero energy. The rate of tunneling out of a superposition then would be regarded as an uncorrelated sum of tunneling rates out of sharp energy eigenstates multiplied by the corresponding weighting factors.

\section*{Acknowledgments}

The author would like to thank Michael Geller, Emily Pritchett, Andrew Sornborger, Fred Wellstood, and Dong Zhou for encouragement and stimulating discussions. He also thanks the anonymous referees for carefully reading the manuscript and for raising many interesting questions.



\begin{thebibliography}{}

\bibitem{LEGGETT80}
A.J. Leggett, Prog. Theor. Phys. (Suppl.) {\bf 69}, 80 (1980)

\bibitem{CALDEIRALEGGETT83}
A.O. Caldeira and A.J. Leggett, Ann. Phys. {\bf 149}, 374 (1983)

\bibitem{CLARKEETAL88}

See, for example, M.H. Devoret, J.M. Martinis, and J. Clarke, Phys. Rev. Lett. {\bf 55}, 1908 (1985); 
J.M.  Martinis, M.H. Devoret, J. Clarke, Phys. Rev. {\bf B35}, 4682 (1987); J. Clarke et al., Science {\bf 239}, 992 (1988); M. R. Geller and A. N. Cleland {\it Phys. Rev.} {\bf A71}, 032311 (2005) and references cited therein.

\bibitem{MAKHLIN01}
For a comprehensive review see Y. Makhlin et al., Rev. Mod. Phys. {\bf 73}, 357 (2001)

\bibitem{COLEMAN-I}
S. Coleman, Phys. Rev. {\bf D15}, 2929 (1977)


\bibitem{BANKSBENDER73}
T. Banks and C. Bender, Phys. Rev. {\bf D8}, 3366 (1973)

\bibitem{JOHNSONETAL2003}
P. Johnson et al.,  Phys. Rev. {\bf B67}, 020509(R) (2003); A. Blais et al., Phys. Rev. Lett. {\bf 90}, 127901 (2003)

\bibitem{GELLERcomment}

I thank Michael Geller for drawing my attention to this important point.

\end{thebibliography}
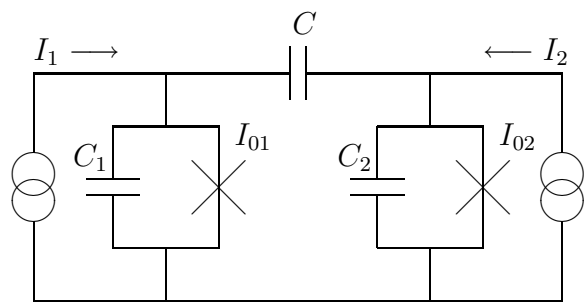
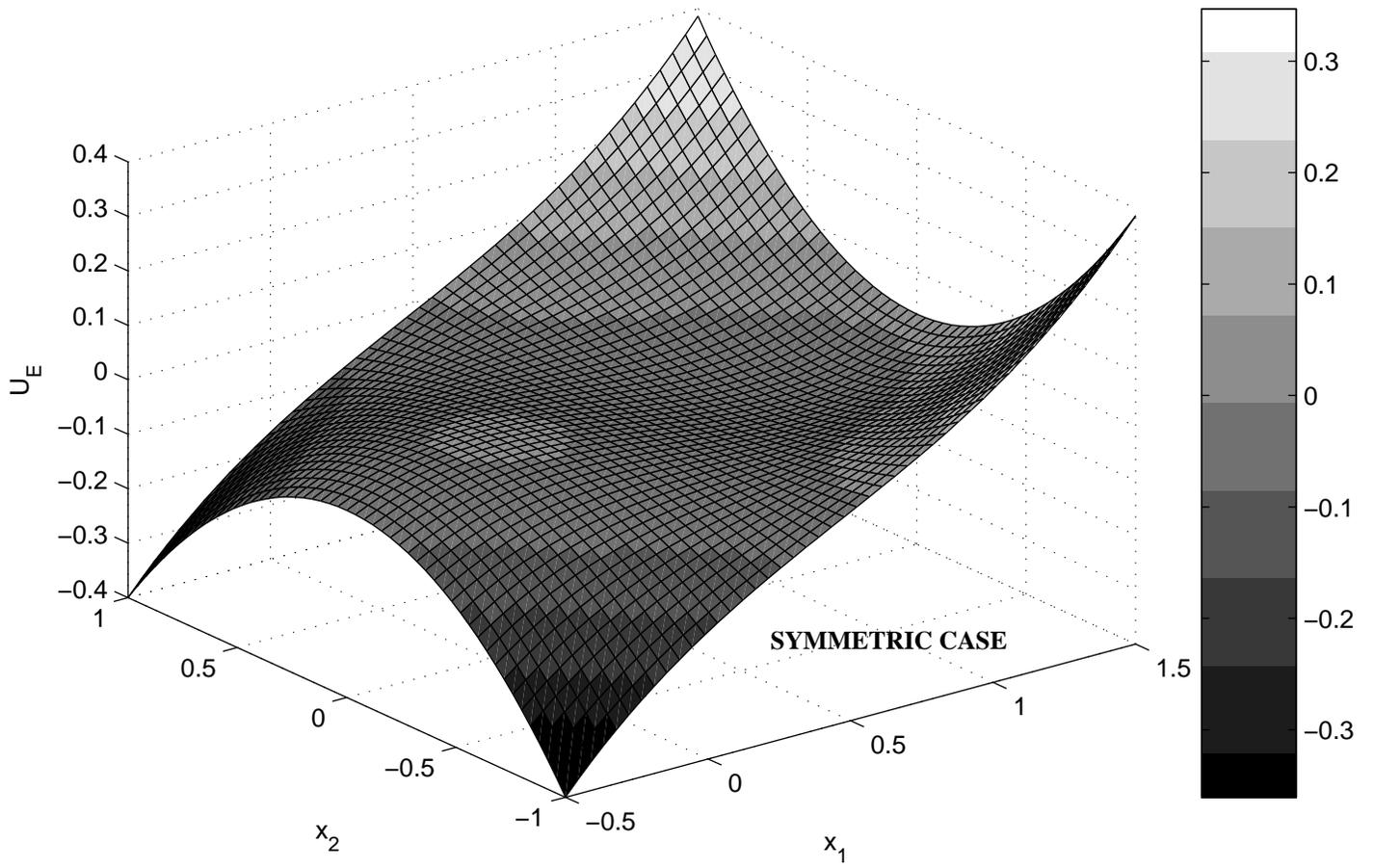
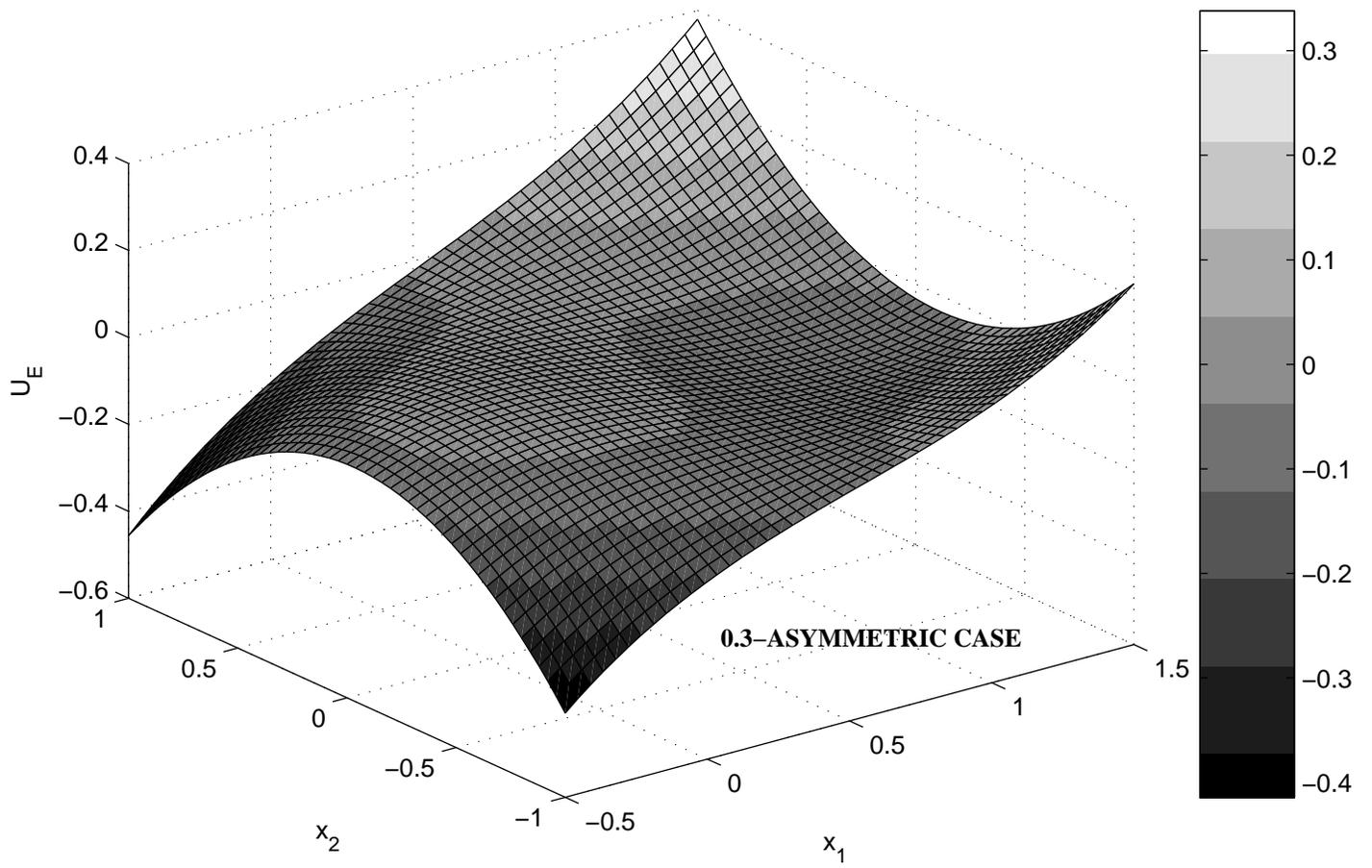
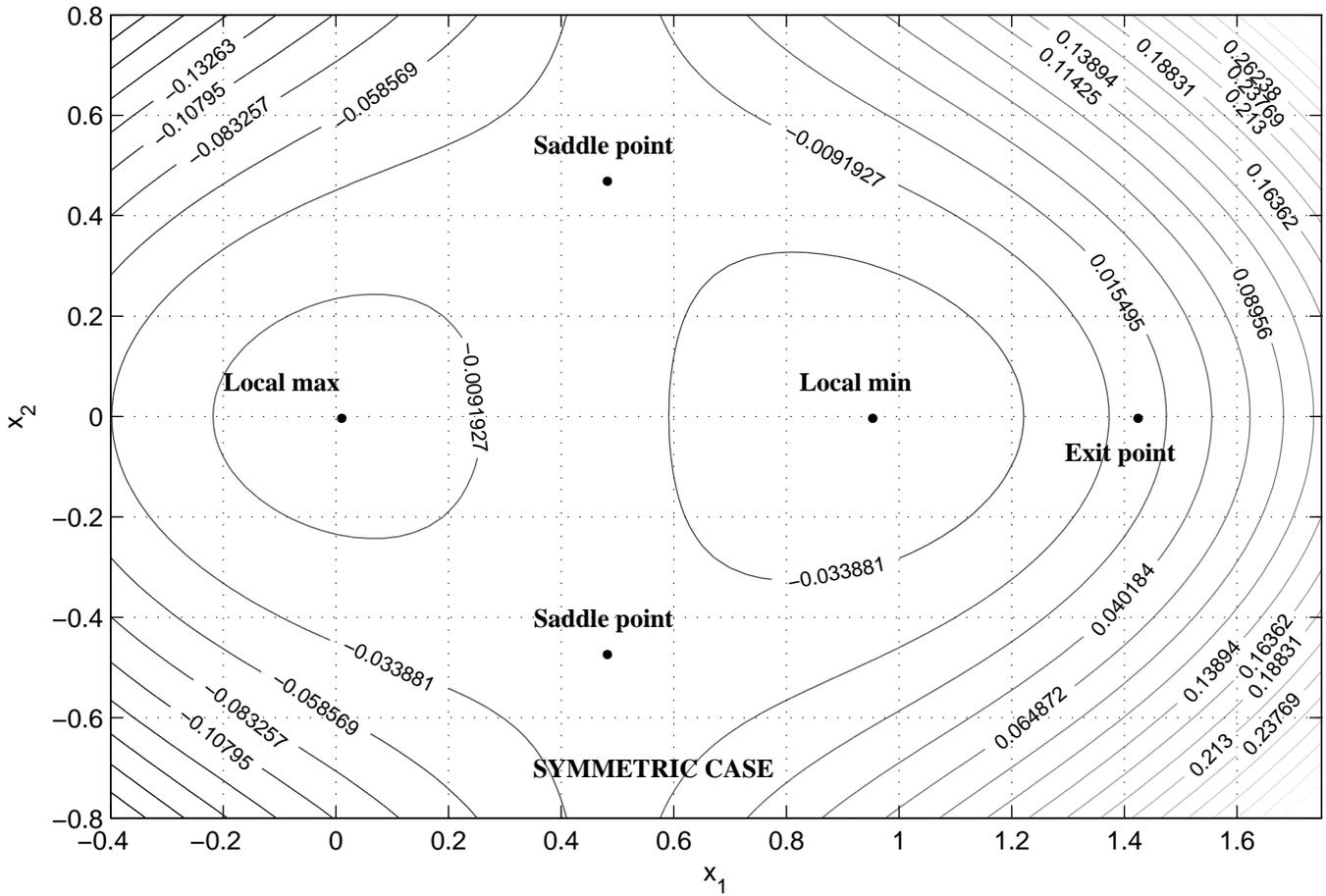
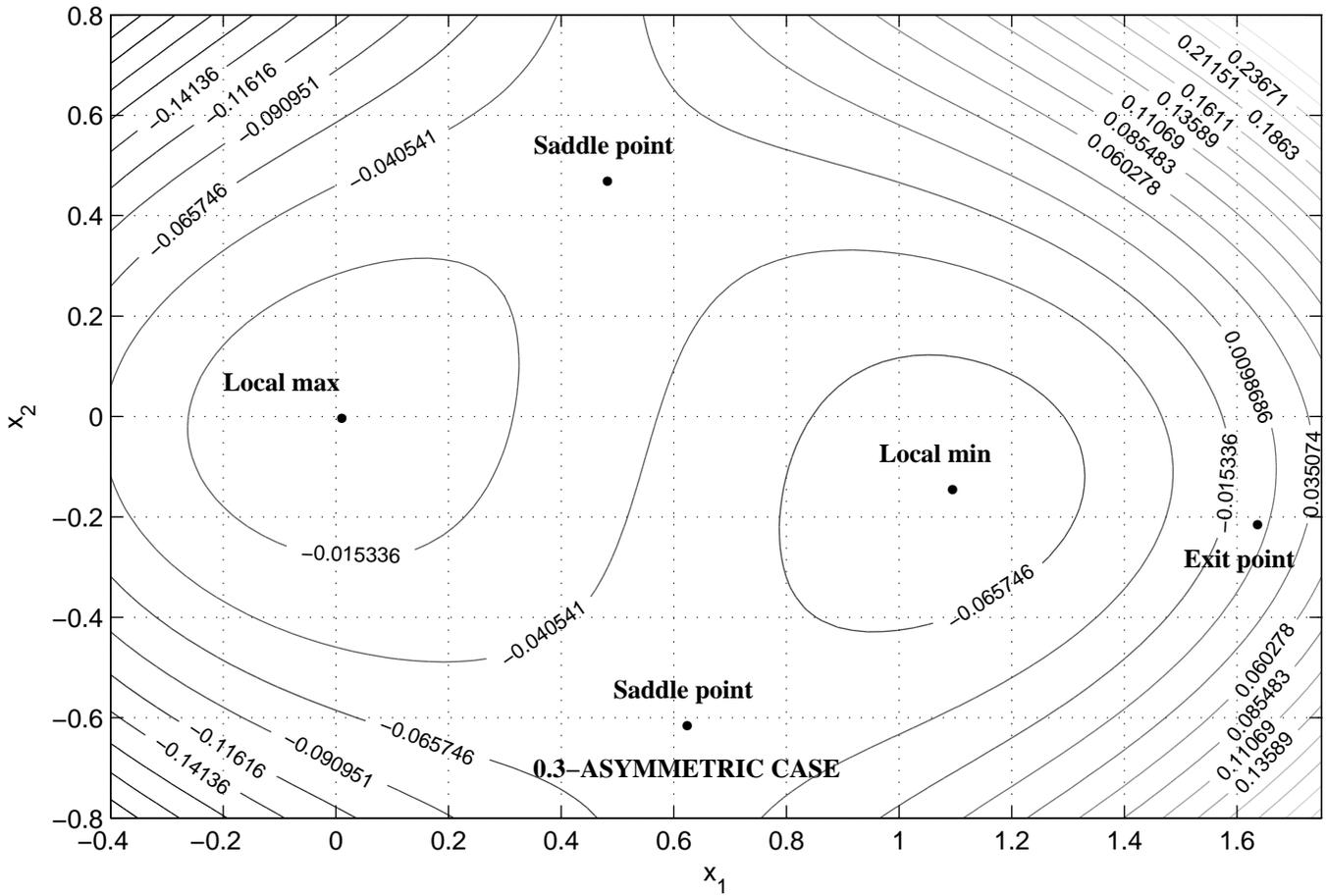
\end{document}